\documentclass[12pt]{article}%
\usepackage{amsmath,latexsym}
\usepackage{graphicx}
\usepackage{amsmath}
\usepackage{amsfonts}
\usepackage{amssymb}%
\begin{document}

\title{{Stable phantom-energy wormholes admitting
     conformal motions }}
   \author{
  Peter K.F. Kuhfittig*\\  \footnote{kuhfitti@msoe.edu}
 \small Department of Mathematics, Milwaukee School of
Engineering,\\
\small Milwaukee, Wisconsin 53202-3109, USA}

\date{}
 \maketitle

\begin{abstract}\noindent
It has been argued that wormholes are as good a prediction
of Einstein's theory as black holes but the theoretical
construction requires a reverse strategy, specifying the
desired geometric properties of the wormhole and leaving
open the determination of the stress-energy tensor.  We
begin by confirming an earlier result by the author
showing that a complete wormhole solution can be
obtained by adopting the equation of state $p=
\omega\rho$ and assuming that the wormhole admits a
one-parameter group of conformal motions.  The main
purpose of this paper is to use the assumption of
conformal symmetry to show that the wormhole is stable
to linearized radial perturbations whenever $-1.5<
\omega <-1$.
\\

\noindent
PAC numbers: 04.20.Jb, 04.20.-q, 04.20.Gz
\end{abstract}

\section{Introduction}\label{S:introduction}

Wormholes are handles or tunnels in spacetime linking widely
separated regions of our own Universe or different universes
altogether \cite{MT88}.  While one could argue that wormholes
are as good a prediction of Einstein's theory as black holes,
the fact remains that a wormhole can only be held open by
violating the null energy condition, which states that the
stress-energy tensor $T_{\alpha\beta}$ must obey the
condition $ T_{\alpha\beta}k^{\alpha}k^{\beta}\ge0 $
for all null vectors \cite{MT88}.  Renewed interest in the
subject of wormhole physics is due in part to the discovery
that our Universe is undergoing an accelerated expansion
\cite{aR98, sP99}, i. e., $\overset{..}{a}>0$ in the Friedmann
equation $\overset{..}{a}/a=-\frac{4\pi}{3}(\rho+3p)$. (Our
units are taken to be those in which $G=c=1$.)  The acceleration
is caused by a negative pressure \emph{dark energy} with equation
of state $p=\omega\rho$, $\omega <-\frac{1}{3}$ and  $\rho>0$.
A value of $\omega <-\frac{1}{3}$ is required for an accelerated
expansion, also referred to as quintessence dark energy.
The value $\omega =-1$ corresponds to the existence of Einstein's
cosmological constant \cite{mC01}.  Of particular interest is
the case $\omega <-1$, usually referred to as \emph{phantom energy},
which is slightly favored over quintessence observationally
\cite{HWY}.  For the phantom-energy case, observe that $\rho+p<0$,
in violation of the null energy condition, thereby satisfying a
fundamental requirement in wormhole physics.  The only real
objection that could be raised is that the notion of dark energy
corresponds to a homogeneous distribution of matter, while
wormhole spacetimes are necessarily inhomogeneous.
Fortunately, the extension to spherically symmetric
inhomogeneous spacetimes has been carried out. (See
Ref.~\cite{sS05} for details.)

When Morris and Thorne \cite{MT88} first proposed
that wormholes may be actual physical objects, they
described the wormhole by the static and spherically
symmetric line element
\begin{equation}\label{E:line1}
ds^{2}=-e^{2\Phi(r)}dt^{2}+\frac{dr^2}{1-b(r)/r}
+r^{2}(d\theta^{2}+\text{sin}^{2}\theta\,
d\phi^{2}).
\end{equation}
Here $\Phi=\Phi(r)$ is called the
\emph{redshift function}, which must be
everywhere finite to avoid an event horizon.
The function $b=b(r)$ is called  the \emph{shape
function} since it helps determine the spatial
shape of the wormhole when viewed, for example,
in an embedding diagram.  The spherical surface
$r=r_0$ is the \emph{throat} of the wormhole and
must satisfy the following conditions: $b(r_0)
=r_0$, $b(r)<r$ for $r>r_0$, and $b'(r_0)<1$,
usually called the \emph{flare-out condition}.
This condition refers to the flaring out of
the embedding diagram pictured in Ref.
\cite{MT88}.  The flare-out condition can
only be satisfied by violating the null
energy condition.  As already noted, in
the present situation, this violation is a
consequence of the phantom-energy background.

The Einstein field equations in the orthonormal
frame, $G_{\hat{\mu}\hat{\nu}}=8\pi
T_{\hat{\mu}\hat{\nu}}$, yield the following
simple interpretation for the components of the
stress-energy tensor: $T_{\hat{t}\hat{t}}=
\rho(r)$, the energy density,
$T_{\hat{r}\hat{r}}=p_r$, the radial pressure,
and $T_{\hat{\theta}\hat{\theta}}=
T_{\hat{\phi}\hat{\phi}}=p_t$, the lateral
pressure.  For the theoretical construction of
the wormhole, Morris and Thorne specified the
functions $\Phi(r)$ and $b(r)$ to obtain the
desired properties of the wormhole, thereby
leaving the components of the stress-energy
tensor dangling.  This strategy would call for
a search for those materials or fields that
yield the required stress-energy tensor.

In a previous paper \cite{pK15}, the author had
addressed this issue by introducing the barotropic
equation of state $p=\omega\rho$, where $\omega <
-1$ is the special case discussed above.  By itself,
this equation of state fails to produce a solution
even if the energy density is known.  The assumption
of conformal symmetry fills the gap in the form of a
complete wormhole solution.

The main purpose of this paper is to use the
assumption of conformal symmetry to obtain conditions
under which the wormhole is stable to linearized
radial perturbations.  The analysis leads to a
physical interpretation of the conformal factor.
We also need to recall briefly the definition of
conformal Killing vectors, as well as the basic
wormhole structure needed to perform the
stability analysis.  That is the topic of the
next section.

A stability analysis of phantom-enery wormholes was
also carried out by Lobo \cite{fL05} but with very
different assumptions, to be discussed later.

\section{Conformal Killing vectors and the shape
function}

In this section we discuss the earlier assumption that
our spacetime admits a one-parameter group of conformal
motions.  Recall that these are motions along which the
metric tensor of a spacetime remains invariant up to a
scale factor.  This is equivalent to stating that there
exists a set of conformal Killing vectors such that
\begin{equation}\label{E:Lie}
   \mathcal{L_{\xi}}g_{\mu\nu}=g_{\eta\nu}\,\xi^{\eta}
   _{\phantom{A};\mu}+g_{\mu\eta}\,\xi^{\eta}_{\phantom{A};
   \nu}=\psi(r)\,g_{\mu\nu},
\end{equation}
where the left-hand side is the Lie derivative of the
metric tensor and $\psi(r)$ is the conformal factor.  The
vector $\xi$ characterizes the conformal symmetry since
the metric tensor $g_{\mu\nu}$ is conformally mapped
into itself along $\xi$.  It is generally agreed that
the assumption of conformal symmetry has proved to be
fruitful in numerous ways, not only leading to new
solutions but also to new geometric and kinematical
insights \cite{HPa, HPb, MM96, MS93, Ray08, fR10, fR12}.

Exact solutions of traversable wormholes admitting
conformal motions  are discussed in Ref. \cite{R2K3}
by assuming a noncommutative-geometry background.
Two earlier studies assumed a \emph{non-static}
conformal symmetry \cite{BHL07, BHL08}.

To discuss the consequences of the
conformal-symmetry assumption, it turns out to
be convenient to use the following form of the
metric:
\begin{equation}\label{E:line2}
   ds^2=- e^{\nu(r)} dt^2+e^{\lambda(r)} dr^2
   +r^2( d\theta^2+\text{sin}^2\theta\, d\phi^2).
\end{equation}
In particular, it is shown in Ref. \cite{pK15} that
\begin{equation} \label{E:gtt}
   e^\nu  =C r^2,
\end{equation}
while
\begin{equation}\label{E:grr}
   e^\lambda  = \psi^{-2}.
\end{equation}
Two of the Einstein field equations are
\begin{equation}\label{E:E1}
  8\pi \rho =\frac{1}{r^2}\left(1 - \psi^2
  \right)-\frac{(\psi^2)'}{r}
\end{equation}
and
\begin{equation}\label{E:E2}
  8\pi p_r=\frac{1}{r^2}\left(3\psi^2-1
  \right).
\end{equation}
[See Ref. \cite{pK15} for details.]

To obtain a wormhole solution, we start with the
equation of state $p_r=\omega\rho$, $\omega <-1$,
discussed in Sec. \ref{S:introduction}, and
substitute Eqs. (\ref{E:E1}) and (\ref{E:E2}):
\begin{equation}\label{E:diff1}
   \frac{1}{8\pi}\frac{1}{r^2}\left(
   3\psi^2-1\right)=\omega\frac{1}
   {8\pi}\left[\frac{1}{r^2}\left(1-
   \psi^2\right)-
   \frac{(\psi^2)'}{r}\right].
\end{equation}
Simplifying, we have
\begin{equation}\label{E:diff2}
   (\psi^2)'+\frac{1}{r}\left(1+\frac{3}{\omega}
   \right)\psi^2=\frac{1}{r}\left(1+\frac{1}{\omega}
   \right).
\end{equation}
This equation is linear in $\psi^2$ and can be
readily solved to obtain

\begin{equation}\label{E:solution}
  \psi^2=\,\frac{\omega +1}{\omega +3}
  +Dr^{-\frac{\omega +3}{\omega}},
\end{equation}
where $D$ is an arbitrary constant.  By
comparing Eqs. (\ref{E:line1}) and
(\ref{E:line2}), we have, in view of Eq.
(\ref{E:grr}),
\begin{equation}\label{E:shape}
   b(r)=r(1-e^{-\lambda})=
     r\left(1-\psi^2\right).
\end{equation}
To satisfy the condition $b(r_0)=r_0$, we must
have $\psi^2(r_0)=0$, which becomes the initial
condition for Eq. (\ref{E:diff2}), thereby
yielding $D$.  The result
is
\begin{equation}\label{E:final}
   \psi^2=\frac{\omega +1}{\omega +3}
   -\frac{\omega +1}{\omega +3}
   r_0^{\frac{\omega +3}{\omega}}
      r^{-\frac{\omega +3}{\omega}}.
\end{equation}
The final forms are
\begin{equation}
  b(r)=r(1-\psi^2(r))
\end{equation}
and
\begin{equation}\label{E:finalfactor}
   \psi^2(r)=\frac{\omega +1}{\omega +3}
   \left(1-r_0^{\frac{\omega +3}{\omega}}
   r^{-\frac{\omega +3}{\omega}}\right).
\end{equation}
Observe that $b(r_0)=r_0$, as noted above.
A simple calculation now shows that since
$\omega <-1$, $b'(r_0)<1$.  So the flare-out
condition is met.

Our final observation is
\begin{equation}\label{E:psi}
   1-\frac{b(r)}{r}=\psi^2(r),
\end{equation}
to be used in Sec. \ref{S:stability}.

\section{Junction to an exterior vacuum solution}

We see from Eq. (\ref{E:gtt}), $e^{\nu}=Cr^2$,
that the wormhole spacetime cannot be asymptotically
flat.  So the wormhole material must be cut off at
some $r=a$ and joined (in the standard way) to an
exterior Schwarzschild solution
\begin{equation}
ds^{2}=-\left(1-\frac{2M}{r}\right)dt^{2}
+\frac{dr^2}{1-2M/r}
+r^{2}(d\theta^{2}+\text{sin}^{2}\theta\,
d\phi^{2}).
\end{equation}
Referring now to line element (\ref{E:line1}),
we first note that $M=\frac{1}{2}b(a)$.  So
for $e^{\nu}=Ca^2$, we have $Ca^2
=1-2M/a$ and the integration constant becomes
\[
   C=\frac{1}{a^2}
   \left(1-\frac{b(a)}{a}\right),
\]
thereby completing the wormhole solution.  The
junction surface plays an important role in the
stability analysis in the next section.

\section{Stability analysis}\label{S:stability}

Our first task in this section is to study the
stresses on the junction surface leading to the
stability criterion.  To that end, let us recall
the Lanczos equations
\cite{fL04}
\begin{equation}\label{E:sigma1}
     \sigma=-\frac{1}{4\pi}\kappa^{\theta}_
     {\phantom{\theta}\theta}
\end{equation}
and
\begin{equation}
   \mathcal{P}=\frac{1}{8\pi}(\kappa^{\tau}
   _{\phantom{\tau}\tau}+\kappa^{\theta}
   _{\phantom{\theta}\theta}),
\end{equation}
where $\kappa_{ij}=K^{+}_{ij}-K^{-}_{ij}$ and
$K_{ij}$ is the extrinsic curvature.  According to
Ref. \cite{fL04},
\begin{equation}
   \kappa^{\theta}_{\phantom{\theta}\theta}=\frac{1}{a}
   \sqrt{1-\frac{2M}{a}}-\frac{1}{a}
      \sqrt{1-\frac{b(a)}{a}}.
\end{equation}
So by Eq. (\ref{E:sigma1}),
\begin{equation}\label{sigma2}
   \sigma=-\frac{1}{4\pi a}
   \left(\sqrt{1-\frac{2M}{a}}-
      \sqrt{1-\frac{b(a)}{a}}\right).
\end{equation}
In view of the assumption $M=\frac{1}{2}b(a)$ in
the previous section, one could reasonably
expect that $\sigma=0$.  However, part of the
junction formalism is to assume that the junction
surface $r=a$ is an extremely thin surface having
a nonzero density.  The mass of the shell is
therefore given by
\begin{equation}\label{E:ms}
  m_s=4\pi a^2\sigma=-a\left(\sqrt{1-\frac{2M}{a}}
  -\sqrt{1-\frac{b(a)}{a}}\right).
\end{equation}
Moreover, given that $\frac{1}{2}b(a)$ is the total
mass inside a sphere of radius $a$, we see that if
$m_s<0$, then $M<\frac{1}{2}b(a)$ and if $m_s>0$,
then $M>\frac{1}{2}b(a)$.

To perform the stability analysis, we make the
usual assumption that the junction surface is a
function of proper time $\tau$ moving about some
equilibrium position $a=a_0$.  Following Lobo
\cite{fL05}, the density takes on the form
\begin{equation}\label{E:sigtau}
   \sigma =-\frac{1}{4\pi a}\left(
   \sqrt{1-\frac{2M}{a}+\dot{a}^2}-
   \sqrt{1-\frac{b(a)}{a}+\dot{a}^2}\right),
\end{equation}
where $\dot{a}=da/d\tau$.

To obtain the stability criterion, one starts
by rearranging Eq. (\ref{E:sigtau}), namely
\[
   \sqrt{1-\frac{2M}{a}+\dot{a}^2}=
   \sqrt{1-\frac{b(a)}{a}+\dot{a}^2}
   -4\pi a\sigma,
\]
to obtain the ``equation of motion"
\begin{equation}
   \dot{a}^2+V(a)=0,
\end{equation}
where $V(a)$ is the potential.  It is a
straightforward exercise to show that
\begin{equation}\label{E:V}
   V(a)=1-\frac{M+b(a)/2}{a}-\frac{m_s^2}{4a^2}-
   \frac{(M-b(a)/2)^2}{m_s^2}.
\end{equation}
The time-dependent radius allows us to study
the effect of a radial perturbation around the
static solution $a=a_0$.  Again following Ref.
\cite{fL05}, this requires linearizing around
$a=a_0$ by considering the Taylor expansion of
$V(a)$ about $a=a_0$:
\begin{equation}
   V(a)=V(a_0)+V'(a_0)(a-a_0)+\frac{1}{2}
   V''(a_0)(a-a_0)^2\\+\text{higher-order terms}.
\end{equation}
To meet the linearized stability criterion, we
must have $V(a_0)=0$ and $V'(a_0)=0$, while
$V''(a_0)>0$.

The question arises in what sense these conditions
are met and how to make best use of Eq. (\ref{E:V}).
To address this issue, consider the last term,
denoted by $L^2$, i. e.,
\begin{equation*}
   L=\frac{M-\frac{1}{2}b(a)}
   {-a\left(\sqrt{1-\frac{2M}{a}}-
      \sqrt{1-\frac{b(a)}{a}}\right)}.
\end{equation*}
After rationalizing the denominator and simplifying,
we obtain
\begin{equation*}
   L=\frac{M-\frac{1}{2}b(a)}{2M-b(a)}
   \left(\sqrt{1-\frac{2M}{a}}+
      \sqrt{1-\frac{b(a)}{a}}\right)\\
      =\frac{1}{2}\left(\sqrt{1-\frac{2M}{a}}
      +\sqrt{1-\frac{b(a)}{a}}\right).
\end{equation*}
Since the equilibrium position $a=a_0$
refers to the junction surface where $2M=b(a)$,
we obtain $L=\sqrt{1-b(a_0)/a_0}$.  The result is
\begin{equation}
   V(a_0)=1-\frac{b(a_0)}{a_0}-\frac{m_s^2}{4a_0^2}
   -\left(\sqrt{1-\frac{b(a_0)}{a_0}}\right)^2,
\end{equation}
which does indeed lead to $V(a_0)=0$ since $m_s=0$.
Now, as noted earlier, part of the junction
formalism is to assume that for a thin shell,
$m_s$ cannot be zero.  So $V(a_0)=0$ must be
viewed as the dividing line between $m_s<0\,$
($2M<b(a_0)$) and $m_s>0\,$ ($2M>b(a_0)$), implying
$V(a_0)$ is equal to zero only in the limit as
$\frac{1}{2}b(a_0)\rightarrow M$.  The same
holds for $V'(a_0)=0$.

The implication is that the dynamic analysis
leading to $V(a_0)=0$ and $V'(a_0)=0$ requires
the use of Eq. (\ref{E:ms}), thereby making
$m_s$ a variable quantity.  But once a
particular junction surface $r=a$ has been
chosen, $m_s$ is necessarily fixed at some
positive or negative value.  So we are
going to make the following assumptions:
to make use of Eq. (\ref{E:V}), we assume
that Eq. (\ref{E:ms}) is no longer needed
and that $m_s^2$ is a small constant.
Given that $b(a)\approx 2M$, these
assumptions lead at once to the
 approximation
\begin{equation}
   V(a)\approx 1-\frac{b(a)}{a}
\end{equation}
for any junction surface $r=a$.  Moreover, the
approximation can be naturally connected to the
conformal symmetry.  In other words, by Eq.
(\ref{E:psi}),
\begin{equation}
   1-\frac{b(a)}{a}=\psi^2(a).
\end{equation}
So $\psi^2$ can be interpreted physically as
an approximation of the potential:
\begin{equation}
   V(a)\approx \psi^2(a).
\end{equation}

From Eq. (\ref{E:finalfactor}), we now get
\begin{equation}
   V''(a_0)=\frac{d^2}{da^2}\psi^2(a_0)=
   -\frac{\omega +1}{\omega}\,\,
   \frac{2\omega +3}{\omega}
   r_0^{\frac{\omega +3}{\omega}}
   a_0^{-\frac{\omega +3}{\omega}-2}.
\end{equation}
Since $\omega <-1$, we conclude that
$V''(a_0)>0$ only if $\omega >-1.5$.  It
follows that our wormhole is stable to
linearized radial perturbations whenever
\[
    -1.5<\omega <-1.
\]
The conclusion is independent of the shape
function and the junction surface.  Lobo's
earlier study \cite{fL05} assumed  a specific
shape function but no conformal symmetry.
Instead, letting $\omega =-2$, a typical
value, the wormhole was found to be stable
for a wide range of values of the radius
of the junction surface.

\section{Conclusion}

For the theoretical construction of a traversable
wormhole, Morris and Thorne proposed the following
reverse strategy: specify the desired geometric
properties of the wormhole, while leaving open the
determination of the stress-energy tensor.  The
first part of this paper confirms an earlier result
\cite{pK15} stating that a complete wormhole solution
can be obtained by adopting the equation of state
$p=\omega\rho$, $\omega <-1$, and assuming that
the wormhole admits a one-parameter group of
conformal motions.  The main purpose of this paper
is to use the assumption of conformal symmetry to
show that the wormhole is stable to linearized
radial perturbations whenever $-1.5<\omega <-1$.
The analysis also yielded a physical interpretation
of the conformal factor in terms of the potential.

\end{document}